\documentclass[utf8]{FrontiersinHarvard} 

\usepackage{url,hyperref,caption,subcaption}
\usepackage[onehalfspacing]{setspace}
\usepackage{graphicx}
\usepackage{bm}
\usepackage{comment}
\usepackage{xcolor}
\usepackage{tabularx}
\usepackage{amsmath}
\usepackage{amssymb}
\usepackage{float}

\def\keyFont{\fontsize{8}{11}\helveticabold }
\def\firstAuthorLast{Heppe {et~al.}} 
\def\Authors{Christian Heppe and Naveen Kumar\,$^{*}$}
\def\Address{Max-Planck-Institut f\"ur Kernphysik, Saupfercheckweg 1, D-69117 Heidelberg, Germany}
\def\corrAuthor{Naveen Kumar}

\def\corrEmail{naveen.kumar@mpi-hd.mpg.de}

\begin{document}
\onecolumn
\firstpage{1}

\title[MeV photons generation in a carbon plasma channel]{High brilliance $\gamma$-rays generation from the laser interaction in a carbon plasma channel} 

\author[\firstAuthorLast ]{\Authors} 
\address{} 
\correspondance{} 

\extraAuth{}

\maketitle

\begin{abstract}

The generation of collimated, high brilliance $\gamma$-ray beams from a structured plasma channel target is studied by means of 2D PIC simulations. Simulation results reveal an
optimum laser pulse pulse duration of $20\,\text{fs}$, for generating $\gamma$-photon beams of brilliances up to $10^{20}\,\text{s}^{-1}\text{mm}^{-1}\text{mrad}^{-2}\,(0.1\,\%\text{BW})^{-1}$ and photon energies well above $200$ MeV in the interaction  of an ultra-intense laser (incident laser power $ P_L \geq 5$ PW) with a high-Z carbon structured plasma target. These results are aimed at employing the upcoming laser facilities with multi-petawatt (PW) laser powers to study the laser-driven nonlinear quantum electrodynamics processes in an all-optical laboratory setup.

\tiny
 \keyFont{ \section{Keywords:} Laser-plasma interaction, Particle-in-cell, $\gamma$-rays generation, Laser-driven QED processes, Radiation generation in plasmas} 
\end{abstract}

\section{Introduction}

Plasma based short-wavelength radiation sources have attracted significant attention in past decades, since plasmas enable not only a compact size but also a wide range of physical mechanisms to generate short-wavelength radiations; be it high-harmonic generation, synchrotron radiation, or betatron radiation mechanisms; ~\cite{Mourou:2006tl,Rousse:2004tc,Kumar:2006wn,Liu:2018va,Yu:2018wb,Ji:2014tv}.  These  highly energetic photon sources have numerous applications for the fundamental research of radiation-reaction force, generation of electron-positron ($e^-e^+$) pairs, photospectroscopy, radiotherapy and radiosurgery;~\cite{Mourou:2006tl}.  The main advantage of using laser-plasma interaction for generating short-wavelength radiation sources is to only require an all-optical setup for the experimental realization. With a continued push for  increasing the laser intensity further in a regime where radiation reaction and pair-production effects become important, the possibility of generating highly-energetic $\gamma$-ray in an all-optical setup is becoming an exciting experimental prospect.

Recently, it has been shown that the use of structured plasma targets \emph{e.g.} a cylindrical target acting as an optical waveguide, is optimum for accelerating electrons and consequently generating $\gamma$-photons;~\cite{Zhu:2015vs,Stark:2016ul,Vranic:2018vz, Jansen:2018tp,Wang:2020uq,Luedtke:2021uq,Huang:2017up}. This scheme is analogous to the betatron radiation generation in an ion channel;~\cite{Rousse:2004tc,Corde:2013tq,Ji:2014tv,Mangles:2005wt,Ji:2018um}. However, here the self-generated magnetic field of the electron beam accelerated in the channel not only causes the generation of $\gamma$ photons but also enhances the yield of these photons, especially in the so-called radiation-dominated regime. Thus, this scheme not only produces higher yields of $\gamma$-photons but also the self-generated magnetic field helps in collimating the generated photon beam in a $10^{\circ}$ lobe around the laser propagation axis. This high-directionality of the photon beams can be exploited for producing electron-positron pairs by the Breit-Wheeler process in colliding two $\gamma$-rays beams setup in a laboratory;~\cite{Wang:2020uq}. Key findings of the scheme are that for high-Z \emph{e.g.} carbon plasmas at incident powers in the range ${P_L}\,\leq 5\,\text{PW}$, the laser to photon energy conversion efficiency drops  for incident laser power in the excess of ${P_L}\,\approx 5\,\text{PW}$;~\cite{Wang:2020uq}. Also, the efficiency of the $\gamma$ photons generation seems to peak around $\tau \sim 45$ fs laser pulse duration for laser powers $P_L \le 10$ PW at laser intensity $I_L= 5\times 10^{22}$ W/cm$^2$;~\cite{Wang:2020uq}. \textcolor{black}{At this laser intensity, radiation reaction can be modeled classically and  stochastic effects involved in quantum radiation reaction are negligible; \cite{Kumar:2013uw}.}

\textcolor{black}{The upcoming laser facilities such as ELI and others;~\cite{Aleonard:2011wy,Papadopoulos:2016wm,xcels,vulcan} are expected to provide multi-petawatt laser systems. These multi-petawatt laser systems are to rely on short laser pulse durations $\tau \sim 20$ fs as significantly increasing the energy contained in the laser pulse is challenging due to technical reasons associated with material damage etc. Thus, it is instructive to examine the generation of $\gamma$-photons with much shorter laser pulses \emph{e.g.} $\tau \le 45$ fs. Also these multi-petawatt laser pulses can be focused to smaller beam radii $\le 10\mu$m resulting in laser intensities ($I_L\ge 1\times 10^{23}$ W/cm$^2$) that can enter the so-called quantum-electrodynamic regime, in which radiation reaction has a stochastic nature and it significantly affects the electron dynamics and consequently $\gamma$-photons generation. Moreover, generation of pair-production can also be important in this regime.
Motivated by these considerations, we study the generation of $\gamma$-photons in a laser-plasma channel, for the laser power exceeding $P_L = 5$ PW. The plasma channel used is a structured carbon plasma target  and the laser pulse has the intensity $I_L= 2.65\times 10^{23}$ W/cm$^2$. Further, we also chose a conical plasma channel to optimize the generation of $\gamma$-photons since conical shaped targets provide higher laser to plasma electron energy conversion efficiencies;~\cite{Vranic:2018vz}. We carry out all simulations for both target geometries for $20\,\text{fs}$ and $40\,\text{fs}$ pulse durations.}

The remainder of this paper is organized as follows: in Sec.\ref{theory} we discuss the simulation setup and  plasma dynamics and the physical process of $\gamma$ photons generation. In sections \ref{cylindrical} and \ref{conical}, we show results from planar and conical plasma channels, respectively. In Sec.\ref{comparison}, we compare our results with previous simulations results. Finally we conclude the discussions in Sec.\ref{conclusions}.

\section{Materials and Methods}\label{results}
\label{theory}

We carry out 2D particle-in-cell (PIC) simulations, employing the open source PIC code SMILEI;~\cite{Derouillat:2018ua}.  The simulation domain is $120\times8\,\mathrm{\mu m}$ ($x \times y$) with a cell size of $0.02\times0.01\,\mathrm{\mu m}$ simulating a time period of $T_{\rm sim}\approx 2500 $ fs, divided into timesteps of $\Delta t \approx 0.02\,\text{fs}$. A linearly polarized laser pulse with wavelength $\lambda_L=0.8\,\mathrm{\mu m}$ impinges on a structured carbon ion plasma target located at $x \geq 10\,\mathrm{\mu m}$ from the left-boundary. We use $16$ particles per cell for electrons as well as ions. \textcolor{black}{To ensure quasi-neutrality in our simulation, the ion  density is chosen to be $n_i=n_e/6$, where $n_e$ is the plasma electron density. Open boundary conditions are used in $x$-direction while periodic boundary conditions are employed in $y$-direction.} The laser pulse has a normalized amplitude $a_0=eE_0/m\omega_0 c=eA_0/m_ec^2 =350$ \textcolor{black}{(corresponding laser intensity $I_L\approx 2.65 \times 10^{23}\, \mathrm{W\, cm^{-2}}$)} and a pulse duration of $\tau = 40\,\text{fs}$ as well as $\tau = 20\,\text{fs}$ (measured at FWHM), where $e$ is the electronic charge, $c$ is the velocity of light in vacuum, $E_0$ ($A_0$) and $\omega_0$ are the laser electric field (vector potential) and frequency, respectively. The core of the plasma channel has density $n_{e,ch}=37 \, n_{cr}$ while the surrounding bulk plasma is denser $n_{e,B}=184 \, n_{cr}$, as also simulated before;~\cite{Stark:2016ul,Jansen:2018tp}. Here $n_{cr}= m_e \omega_0 / 4 \pi e^2$ is the non-relativistic critical plasma density.  \textcolor{black}{This type of plasma channel can either be created using modern techniques; see~\cite{Fischer:2013aa} or they can arise dynamically due to the action of ponderomotive force associated with the laser pre-pulse.} The laser pulse has a 2D-Gaussian spatial distribution and it was focused on the center of the channel's opening at $x=10 \mathrm{\mu m}$ and $y=4 \mathrm{\mu m}$ from the left-boundary of the simulation box. To maximize the energy conversion from the laser pulse to plasma electrons, the waist of the pulse $w_0$ in the focal plane was chosen to be equal to the channels entrance radius $w_0=R_0$. 
On increasing the laser waist-radius, one can scan the power dependence $P_L=\pi w_0^2 I_0/2=\pi R_0^2 I_0/2$, in our simulations, where $I_0$ is the peak laser intensity. The first set of simulations was carried out for a planar target that represents a longitudinal cross-section of a cylindrical target with a constant channel radius $R(x)=R_0$. For a conical target the radius varies as $R(x)=R_0-({R_0-R_{\rm exit}})(x-10\mu m)/{L}$,  for $x \geq 10 \mathrm{\mu m}$, where $L=110\mathrm{\mu m}$ is the channel length and $R_0$ and $R_{\rm exit} = 0.25\mu m$ for all incident laser powers. In total the experiment consists of two sets of four simulations. We scanned the incident power for $P_L=[5,10,15, 20]\,\text{PW}$ for laser pulse durations of $\tau = 20$ and $40 \,\text{fs}$.

The laser pulse parameters chosen to be broadly consistent with the upcoming laser systems at ELI facility, which aim to investigate laser-driven quantum-electrodynamic processes. \textcolor{black}{SMILEI employs a fully stochastic quantum Monte-Carlo model of photon emission and pair generation by the Breit-Wheeler process; see ~\cite{breittable,radtable}. The probability of photon generation and pair-creation can be simplified considerably if some assumptions can be enforced, \emph{e.g.} ultra-relativistic particle motion, the electromagnetic field experienced by particles in their rest frames are lower than the critical Schwinger field and varies slowly over the formation time of a photon, and radiation emission by particle is incoherent;~\cite{Nikishov:1964aa}. These assumptions are always satisfied in the PIC simulations carried out here.  The photon-emission and pair-creation are fundamentally a random-walk process; \cite{Duclous:2010aa,Kirk:2009aa,breittable,radtable}.  One assigns initial and final optical depths (between $0$ and $1$) to a photon. This optical depth is allowed to evolve in time following particles motion in the laser field. The time evolution of the optical depth is  equal to the production rate of pairs in the laser field;~\cite{breittable}. When the final optical depth is reached, a photon is allowed to be emitted by the algorithm. The parameters of emitted particles can be obtained by inverting the cumulative probability distribution function of the respective species; see ~\cite{Duclous:2010aa,Lobet:2015ve}. Production of $e^-e^+$ pairs from photons also utilizes a similar procedure and pairs are expected to be  emitted along the photon propagation direction; see ~\cite{breittable}.}

\subsection{Filamentation  of the laser pulse in a plasma channel}

\textcolor{black}{As one increases the incident laser power at a fixed laser intensity, the focal spot of the laser pulse increases. For high laser power (and large laser spot-size), the laser pulse becomes susceptible to the laser filamentation instability;~\cite{Kaw:1973aa,Sheng:2001aa,Kumar:2006uy}.
This issue hitherto has not been discussed in the previous studies so far, even though the filamentary structures in electron plasma density are visible and they are attributed to the current filamentation instabilities;~\cite{Jansen:2018tp}. Transverse laser pulse filamentation can also affect the generation of $\gamma$-photons in a plasma channel. Thus, it is instructive to estimate the laser filament and choose the laser spot size which is smaller than the filament size due to the filamentation instability.}
For the purpose of estimating the filament size of a laser pulse in an underdense plasma, we use the well-know formalism of laser-driven parametric instabilities and employ the envelope model of the laser pulse propagation. For including the radiation reaction force on the instability analysis, we follow the approach developed in \cite{Kumar:2013uw} by including the dominant term of the Landau-Lifshitz radiation reaction force.
Equation of motion for an electron in the laser electric and magnetic fields including the leading order term of the Landau-Lifshitz radiation reaction force is 
\begin{equation}
\frac{\partial \bm{p}}{\partial t} + \bm{\upsilon}\cdot \nabla\bm{p}=-e \left(\bm{E}+\frac{1}{c}\bm{\upsilon}\times\bm{B}\right) -\frac{2e^4}{3m_e^2c^5}\gamma^2\bm{\upsilon}\left[\left(\bm{E}+\frac{1}{c}\bm{\upsilon}\times\bm{B}\right)^2-\left(\frac{\bm{\upsilon}}{c}\cdot \bm{E}\right)^2\right],
\label{eom}
\end{equation} 
where $\gamma=1/\sqrt{1-\bm{\upsilon}^2},\, e$ is the electronic charge, $m_e$ is the electron mass, and $c$ is the velocity of light in vacuum. The other terms of the Landau-Lifshitz radiation force are $1/\gamma$ times smaller than leading order term;\cite {Landau:2005fr}.

For theoretical calculations, we employ the circularly polarized laser pulse propagating in a plasma. \textcolor{black}{The relativistic motion of an electron in a linearly polarized laser pulse involves generation of high-harmonics at the fundamental laser frequency. Due to this reason, the Lorentz factor $\gamma$ of an electron is not constant in time and analytical treatment of any laser-driven plasma processes becomes intractable in ultra-relativistic regime. For circularly polarized laser pulse, the gamma factor is constant in time and it enables analytically tractable results to showcase the influence of radiation reaction on the filamentation instability of a laser pulse in a plasma. This has been also done by others in the past while investigating the parametric instabilities of laser pulse in plasmas. A quick comparison with the linearly polarized laser pulse can be made by rescaling the normalized vector potential $a_0$ as $a_0^{\rm LP}=a_0^{\rm CP}/\sqrt{2}$. We express the electric and magnetic fields in potentials employing Coulomb gauge in Eq.\eqref{eom}, and write the CP laser pulse as $\bm{A}=\bm{A_{0}}(\bm{x}_{\perp},z,t)e^{i\psi_0}/2+c.c.$, where $\psi_0=k_0 z-\omega_0t$}. We have assumed that the laser pulse amplitude varies slowly i.e. $ \left|\partial \bm{A}_{0}/{\partial t}\right|\ll\left|\omega_0 \bm{A}_{0}\right|,\, \left|\partial \bm{A}_{0}/{\partial z}\right|\ll\left|k_0 \bm{A}_{0}\right|$, and $|\phi| \ll |\bm{A}|, \omega_p^2/\gamma \omega_0^2 \ll 1$, and $\gamma=(1+e^2|\bm{A}|^2/m_e^2c^4)^{1/2}$, $\phi$ being the electrostatic potential. We then write the transverse component of the quiver momentum from Eq.\eqref{eom} as
\begin{gather}
\frac{\partial}{\partial t}\left(\bm{p}_{\perp}-\frac{e}{c}\bm{A}\right)=-\frac{e\mu\omega_0}{c}\bm{A}\gamma |\bm{A}|^2,
\label{qmom}
\end{gather}
where  $\omega_p=(4 \pi n_e e^2/m_e)^{1/2}$ the non-relativistic plasma frequency, and $\mu=2e^4\omega_0/3m_e^3c^7$. We have assumed $|\mu \gamma |\bm{A}|^2| \ll 1$, which is valid for laser intensities $I_{L}\le 10^{23}\,\text{W/cm}^2$, for which the influence of radiation reaction force has to be taken into account. The wave equation then reads as
\begin{equation}
\nabla^2\bm{A}-\frac{1}{c^2} \frac{\partial^2 \bm{A}}{\partial t^2}=\frac{\omega_p^{2}}{\gamma c^2}\frac{c}{e}\bm{p}_{\perp},
\label{waveeq}
\end{equation}
where $A_0$ is the amplitude of the envelope.
On collecting the terms containing $e^{i\psi_0}$, Eq.\eqref{waveeq} yields the dispersion relation for the equilibrium vector potential as $\omega_0^2=k_0^2c^2+{\omega_p^{'2}}\left(1-{i\mu}|A_0|^2\gamma_0/2\right)$, where $\gamma_0=(1+e^2 A_0^2/2 m_e^2 c^4)^{1/2}$ is the equilibrium Lorentz-factor, \textcolor{black}{and $\omega_p^{'}=(4 \pi n_e e^2/m_e \gamma_0)^{1/2}$ is the relativistic electron plasma frequency corresponding to the equilibrium propagation of the laser pump}. It is evident from the dispersion relation that the radiation reaction term causes damping of the pump laser field. This damping can be incorporated either by defining a frequency or a wavenumber shift in the pump laser by defining a frequency shift of the form $\omega_0=\omega_{0r}-i\Delta\omega_0,\,\Delta\omega_0 \ll \omega_{0r}$(real part of $\omega_0$) with the frequency shift $\Delta\omega_0$ being $\Delta\omega_0={\omega_{p}^{'2}\varepsilon\gamma_0 a_0^2}/{2\omega_{0r}}$, where $\varepsilon=r_e\omega_{0r}/3c,$ and $r_e=e^2/m_e c^2$ is the classical radius of the electron. Eq.\eqref{waveeq} in the envelope approximation can be expanded as
\noindent
\begin{equation}
	2i\omega_0 \frac{\partial A_0}{\partial t} + c^2 \nabla_\perp^2 A_0+ \frac{\omega_p^2}{\gamma_0}\left(1-\frac{{i\mu}|A_0|^2\gamma_0}{2}\right)A_0  = \frac{\omega_p^2}{\gamma}\left(1-\frac{{i\mu}|A|^2\gamma}{2}\right)A,
	\label{eq4}
\end{equation}
\textcolor{black}{The left hand side of Eq.\eqref{eq4} represent the equilibrium propagation of the laser pulse in an envelope approximation. While the right hand side of Eq.\eqref{eq4} is the source of perturbation for the filamentation instability.} Following the approach in \cite{Kumar:2006uy}, we write $\bm{A}_{0}(\bm{x}_{\perp},z,t)=\bm{A_{0}}+ \bm{ \delta A}(\bm{x}_{\perp},z,t)$, and $\bm{ \delta A}=\bm{ \delta A}_r+i \bm {\delta A}_i$, $\bm{ \delta A}_r,\,\bm {\delta A}_i \sim {\rm exp}(i {\bm q}_{\perp}{\bm x}_{\perp}-i\delta \omega t)$. This yields two equations for real and imaginary parts of perturbation amplitudes $\bm {\delta A}_r$ and $\bm {\delta A}_i$, yielding the growth rate $\Gamma = \text{Im}(\delta\omega)-\Delta\omega_0$, as
\noindent
\begin{equation}
	\Gamma \approx \frac{{q}_\perp c^2}{4 \omega_0^2}\left[\frac{\omega_p^2a_0^2}{2 \gamma_0^3}-q_\perp^2c^2\right]^{1/2}-\frac{\varepsilon a_0^2\omega_p^2}{\omega_0}.
	\label{Filaments}
\end{equation}
From here the reduction in the filamentation growth rate is apparent. Thus, radiation reaction unlike in the case of stimulated Raman scattering;~\cite{Kumar:2013uw}, does not enhance the growth of the filamentaion instability. \textcolor{black}{This is not unexpected since the enhancement in the case of stimulated Raman scattering depends on the simultaneous resonant excitations of Stokes and anti-Stokes modes in the plasma. Radiation reaction force causes mixing of these modes, leading to higher growth rate of the stimulated Raman scattering. The spatial filamentaion instability, as discussed here, is a broadband instability since the growth can occur over a large range of frequencies. Since the mixing of two distinct modes is absent for filamentation instability, radiation reaction does not lead to the enhanced growth rate of filamentation instability but plays the role of a damping force in a plasma.} From this equation we find the filament size to be $q_\perp^{-1}\approx (c/\sqrt{2}\omega_p)a_0$. For our setup (linearly polarized laser pulse) with $a_0=350$ and \textcolor{black}{$n_{e, ch}=37 n_{cr}$}, this yields an approximative size of $q_\perp^{-1} \sim 7.3\,\mathrm{\mu m}$. We expect strong filamentation for a spot-size or channel width exceeding $q_\perp^{-1}$, resulting in the loss of efficiency in high-energy photon production. Simulation setup in all cases  (Also in \cite{Heppe:2020ub}) have spot-size smaller than optimum filament size given by the above scaling to improve the photon generation especially for the pulse duration $\tau = 20\,\text{fs}$.

\section{Results}

First we qualitatively discuss and recapitulate the plasma dynamics involved in the generation of the ultra-strong magnetic field and the high-energy photon emission. Afterwards we show results on the photon beam properties from a 2D planar target and afterwards from a conical target. The plasma physical processes involved remain qualitatively true for both targets.

\subsection{Plasma dynamics and MeV photon emission}

The general features of the plasma dynamics involved in the generation of $\gamma$-photons qualitatively show similar  behavior as discussed before;~\cite{Stark:2016ul,Wang:2020uq,Jansen:2018tp,Heppe:2020ub}. 
The laser pulse can accelerate plasma electrons to very high energy in this plasma channel via direct laser acceleration;~\cite{Vranic:2018vz}. \textcolor{black}{The current associated with these so-called hot electrons often exceed the so-called Alfv\'enic current. Consequently, a return plasma current is excited which compensates for the hot electron current and enables their transportation inside the plasma. The magnetic field $B_z$ associated with the hot-electrons can help in generating energetic photons and enables a large laser energy conversion into $\gamma$-photons. If these accelerated hot-electrons have transverse dimensions of  $\sim c/\omega_{\rm p}$, where $\omega_{\rm p}$ is the background plasma frequency corresponding to the surrounding bulk plasma, then a filamentation by the counter-propagating background plasma current also ensues;~\cite{Kumar:2009te}. Since the return plasma current filamnetation is associated with the bulk plasma density, the generated quasi-static magnetic field due to the filamentation instability exceeds the magnetic field generated by the forward moving relativistic electrons in the plasma channel and dominates the generation of high-energy $\gamma$ photons by synchrotron emission mechanism;~\cite{Stark:2016ul,Jansen:2018tp}. } This radiated synchrotron power $P_{\textrm{rad}}$ is proportional  to the emissivity parameter $\eta$ \emph{i.e.}$P_{\textrm{rad}} \propto \eta^2$. This parameter $\eta$ reads as
\begin{equation}
	\label{PowSynch}
		\eta \equiv \frac{\gamma}{E_s} \sqrt{\left(\mathbf{E}+\frac{1}{c}\left(\mathbf{v}\times\mathbf{B}\right)\right)^2-\frac{1}{c^2}\left(\mathbf{E}\cdot\mathbf{v}\right)^2},
\end{equation}
where $\gamma$ is the Lorentz factor for electron, $\mathbf{v}$ its velocity, $\mathbf{E}$ and $\mathbf{B}$ are local electric and magnetic fields, and $E_s\approx1.3\times10^{18}\,\text{V} \text{m}^{-1}$ is the so called \textit{Schwinger field}. \textcolor{black}{If one were to only consider the laser electric ($E_L$) and magnetic ($B_L$) fields in Eq.\eqref{PowSynch}, the parameter $\eta$ would be close to zero on account of the Doppler shifted electric field experienced by the hot-electrons, in their rest frame, co-propagating with the laser. However, the plasma magnetic field ($B_p$) generated due to hot-electrons and dominantly by the return current filamentation can facilitate non-zero value of the parameter $\eta$ since now $B=B_L+B_p$ and despite the Doppler shift cancellation of laser electric and magnetic fields, one still has $B_p$ facilitating non-zero value of $\eta$.  Consequently, MeV photon generation in a plasma channel can ensue. One may also note that pair-production by these $\gamma$-photons in the presence of ultra-intense magnetic field can also occur. For the parameters considered here, we see negligible pair-production in our simulations. This is in sync with previous PIC simulations.} Fig.\ref{fig:1} (see also the movie in Supplemental Data) shows the electron density $n_e$, the azimuthal magnetic field $B_z$, and a composite figure of $n_e$ (grey) overlayed by the synchrotron emissivity factor $\eta \geq 0.0004$ (red), shortly after the laser pulse hits the target.  Due to the ponderomotive force, there is an accumulation of the plasma electron density at the boundary of the channel;~\cite{Huang:2017up}.  The fluctuations associated with the plasma channel boundaries are presumably due to hosing type instabilities associated with the laser pulse propagation in a plasma channel.
The plasma electrons become relativistic soon due to direct laser acceleration:~\cite{Huang:2017up, Jirka:2020we, Vranic:2018vz}.
 In the centre of the plasma channel, a strong forward current associated with the electrons is generated;~\cite{Vranic:2018vz}. These electrons generates a strong and quasi-static azimuthal magnetic field $B_z$ up to the order of $B_0 \sim 4\mbox{-}7\, \text{MT}$ as it propagates along the targets symmetry axis. As these electrons propagate through the channel, they excite a return plasma current along the plasma channel boundary. As discussed before, the filamentation of the current also causes the generation of quasi-static ultra-strong magnetic field. Since this magnetic field is caused by the Weibel type filamentation instabilities, it is quasi-static (owing to the Weibel-type instability being aperiodic in a collisionless plasma; see ~\cite{Kumar:2009te}) and does not propagate deeper into the plasma, as seen in Fig.\ref{fig:1b} (see also the Supplemental Data). This can be verified from Fig.\ref{fig:Eta} where  the strongest radiation emission is shown to occur in the overlapping area of magnetic field and high $\eta$ factor closer to the target surface. 
  
The high-energy photon emission shows a broader angular distribution, albeit with the presence of two well defined peaks situated around $\pm 45^{\circ}$ from the laser propagation direction in the channel as visible in Fig.\ref{fig:Angular}.  From Fig.\ref{fig:Angular}, we can see that high-energy photons ($\gtrsim 50 \,\text{MeV}$) are concentrated in these two peaks. This occurrence of two peaks spectra has also been noted before for the case of linearly polarized laser pulse propagation in plasmas;~\cite{Jansen:2018tp, Xue:2020tz}. \textcolor{black}{The physical reason for two-lobes in the case of linearly polarized laser pulse stems from the fact that the laser field attains absolute maximum twice in a laser cycle, causing significant electron heating twice in the same laser cycle. This results in two-distinct lobes of radiation emission. While in the case of circularly polarized laser field, the maximum absolute electric field remains constant in a laser cycle and one only sees a single lobe of emitted radiation;~\cite{Luedtke:2021uq}.} From now onwards, while discussing the properties of the  $\gamma$-photon beams, we only consider photons emitted within one emission lobe at $\Phi = 45^{\circ}\pm 10^{\circ}$.
 
\textcolor{black}{Fig.\ref{fig:tau-scan} shows fractional efficiency of laser to photon energy conversion ($\varepsilon$) with the laser pulse duration for the incident laser power $P_L=10$ PW for the planar target. The choice of $P_L=10$ PW becomes clear later while comparing the brilliance of the photon beams. As evident the efficiency peaks for the laser pulse duration $\tau=20$ fs. This is attributed to plasma instabilities associated with the laser pulse propagation not being dominant for the short laser pulse duration of $\tau=20$ fs. However, for longer laser pulse duration, one can expect the generation of higher energy $\gamma$-photons. Thus, from now onwards we show the $\gamma$-photons spectra for two laser pulse durations \emph{e.g.} $\tau=20$ and $\tau=40$ fs. }

\subsection{Planar target}\label{cylindrical}

Fig.\ref{fig:Spectrum-cylindric} shows the photon spectra generated for the different laser powers and pulse lengths. The maximum energy emitted by a photon increases with the incident laser power with maximum photon energy emitted $\ge 250$ MeV for $P_L=10$ PW laser power. The emitted photon spectra do not show significant deviations with laser power at different pulse lengths.

To quantify the quality of the emissions regarding the collimated beams we take a look at the two distinct radiation lobes as seen in Fig.\ref{fig:Angular}. Here we consider all photons emitted with energies $\geq 1\,\text{MeV}$ within the angular range described before, \emph{i.e.} only considering one of the two photon beams emitted. Tables \ref{tab:cylindrical20} and \ref{tab:cylindrical40}
give a summary of the brilliance of this single photon beam  as well as the laser to collimated $\gamma$-ray photon beam ($\epsilon$) energy conversion efficiencies for different laser powers. For a $\gamma$-ray photon beam with a cut-off of $1$ MeV photon energy, the recorded brilliances are higher for the laser pulse duration $\tau=20$ fs \textcolor{black}{compared to pulse duration $\tau=40$ fs, except for $P_L=10$ PW case (see Tables \ref{tab:cylindrical20} and \ref{tab:cylindrical40}).} \textcolor{black}{This observation is particularly important. On one hand the use of a longer laser pulse duration can enhance the highest $\gamma$-photon emitted energy as seen in Figs.\ref{fig:Spectrum-cylindric} and \ref{fig:Spectrum-conical}. Moreover, the brilliances for $10$ MeV cut-off photon energies are also better for the shorter laser pulse duration $\tau=20$ fs at all incident laser powers except for $P_L=10$ PW.   The lower brilliance for $\tau=40$ fs laser pulse duration is on account of these $\gamma$-photons being not as tightly collimated in the angular range $\Delta\theta=20^{\circ}$. This clearly points to deleterious role of other plasma instabilities such as hosing instability on the radiation emission. Since the hosing instability can cause laser focus to jitter along the equilibrium propagation direction, resulting in a wide angular radiation emission spectrum. These plasma instabilities associated with the laser propagation are dominant for longer laser pulse durations, and consequentially strongly affect the brilliances for longer pulse duration case.} The brilliance of the photon beams seems to be weakly dependent on $\epsilon$. This is presumably due to higher abundance of low-energy ($\epsilon_{\gamma} \ll$ 1 MeV) photons in the lobe. A similar trend is also noted for longer laser pulse duration \emph{e.g.} $\tau =40$ fs (see Table \ref{tab:cylindrical40}).  \textcolor{black}{The brilliances for 10 MeV photons beam are considerably lower than for 1 MeV photon beams. This is attributed to the following two reasons: first relatively low numbers of higher $\gamma$-photons ($\ge 10$MeV) generation compared to 1 MeV photons, and second the spatial distribution of these 10 MeV photons not confined to a small angular range used to record the brilliance. One can also see from Tables \ref{tab:cylindrical20} and \ref{tab:cylindrical40}, that the normalized photon counts $\tilde{N}_{\gamma,\rm max}$ normalized to the photon count from the $P_L=5$ PW case, first shows higher photon yields and then start saturating. However, the brilliances of both 1 MeV and 10 MeV $\gamma$-photons peak at $P_L=20$ PW, hinting that $\gamma$-photons are largely generated in a small angular range ($\Delta \theta = \pm 10^{\circ}$ around one lobe).} The highest $\gamma$-ray photon brilliance recoded \emph{i.e.} $\sim 4.00\times10^{20}\,\text{s}^{-1}\text{mm}^{-1}\text{mrad}^{-2}\,(0.1\,\%\text{BW})^{-1}$ for photon energy cut-off $1$ MeV and $\sim4.38\times10^{18}\,\text{s}^{-1}\text{mm}^{-1}\text{mrad}^{-2}\,(0.1\,\%\text{BW})^{-1}$ for photon energy cut-off $10$ MeV  are sufficient for the observation of pair-generation due to linear Breit-Wheeler and photon-photon scattering processes envisaged in the upcoming projects;~\cite{Abramowicz:2021uf,He:2021va,Kettle:2021ti}.

\subsection{Conical target}\label{conical}

Fig.\ref{fig:Spectrum-conical} shows the photon energy spectra for a conical target.  The total number of photons is of the same order of magnitude as for the planar target around $\sim10^{33} \,\text{MeV}^{-1}$. Tables \ref{tab:conical20} and  \ref{tab:conical40} summarize the brilliance for $\gamma$-ray beams generated for both $1$ and $10$ MeV cut-off energies. The conical target, for the laser pulse duration $\tau=20$ fs, does show higher fractions conversion efficiency of the laser energy to the collimated photon beam.
However, this higher fraction does not yield a comparable higher photon brilliance compared to the planar target case, as seen in from Table \ref{tab:conical20}. This is again attributed to generation of photons with energies lower than $\ll 1$ MeV, \textcolor{black}{which are not shown in Fig.\ref{fig:Spectrum-conical}}. However, as in the case of a planar target, one sees the optimum laser pulse duration appears to be $\tau=20$ fs for  generating $\gamma$-ray beams of highest brilliance for all incident laser powers and both cut-off energies. The highest brilliance in this case is slightly lower than the planar target case, see Table \ref{tab:cylindrical20}. The enhanced laser energy coupling to electrons in the conical target case, see \cite{Yu:2012ww}, does not yield comparable higher photon brilliances. \textcolor{black}{The normalized photon count $\tilde{N}_{\gamma,\rm max}$, as shown in Tables \ref{tab:conical20} and  \ref{tab:conical40}, show a linear increase in photon counts with higher laser power but the brilliance of the $\gamma$-photons does not drastically improve at higher laser power $P_L=20$ PW. This suggest that though an efficient laser energy conversion to photons occurs, the laser pulse is subjected to strong jitter, presumably associated with hosing type instabilities, causing the $\gamma$-photons to be generated in large angular range thereby reducing the brilliance of the $\gamma$-photons beam.}

\subsection{Comparison}\label{comparison}
Our results show similar trends as observed before for heavier plasma ions. For a more intense $a_0=468$, $\lambda=1\, \mathrm{\mu m}$ pulse with a $15\,\text{fs}$ duration using a structured target made of hydrogen plasma and pre-ionized gold, \cite{Gu:2018uf} achieved brilliance levels of $\sim 10^{24}\, \,\text{s}^{-1}\text{mm}^{-1}\text{mrad}^{-2}\,(0.1\,\%\text{BW})^{-1}$ at photon energies of $58\, \text{MeV}$ and generation of photons with maxmimum energies up to $1.5\, \text{GeV}$. However the energy conversion in their case lies around $1.45\%$, which is considerably lower than the highest energy conversion we have observed in our simulations, especially for the conical target viz. $4\% \lesssim \epsilon_{beam} \lesssim 10\%$ using the $20 \,\text{fs}$ pulse (see Table \ref{tab:conical20}). Similarly \cite{Xue:2020tz} showed that for $a_0=150$, $\lambda=1\, \mathrm{\mu m}$ pulse of duration $\tau\approx40$ fs, brilliance levels of $\sim 10^{21}\,\,\text{s}^{-1}\text{mm}^{-1}\text{mrad}^{-2}\,(0.1\,\%\text{BW})^{-1}$ at $1\,\text{MeV}$, can be achieved for a similarly structured target consisting of hydrogen plasma surrounded by Al$^{3+}$ bulk plasma. We observe brilliances of an order of magnitude lower but this is comparable to the results shown in Fig.\ref{fig:Thesisefficiencybeam-log} for hydrogen targets; see~\cite{Heppe:2020ub}. In a second set of simulations shown by \cite{Xue:2020tz}, consisting of Au-cone filled with a hydrogen plasma produced a two-lobe $\gamma$-ray beam with energies up to $\le 420\,\text{MeV}$ with brilliances of again $\sim 10^{21} \text{s}^{-1} \text{mm}^{-1} \text{mrad}^{-2} (0.1\,\%\text{BW})^{-1}$ at $1$ and $10\,\text{MeV}$ photon cut-off energies.  Also \cite{Wang:2021tp} showed for $a_0=100$, $\lambda=1\, \mathrm{\mu m}$, pulse duration $\tau=30\,\text{fs}$ and a Au$^{+69}$-plasma with electron density $n_e=276 n_{cr}$, generations of photons with energy up to $\sim 1.5\,\text{GeV}$.  The generated photon beams had a brilliance of $2.9\times10^{21}\,\,\text{s}^{-1}\text{mm}^{-1}\text{mrad}^{-2}\,(0.1\,\%\text{BW})^{-1}$ at $1$ MeV, which is very similar to our results. In our simulations carried out before \cite{Heppe:2020ub} with similar laser settings ($a_0=190$, $\lambda=0.8\, \mathrm{\mu m}$ and $\tau=40\,\text{fs}$) using the same 2D-cylindrical target but consisting of hydrogen-plasma yielded the same two-lobe distribution we see in Fig.  \ref{fig:Angular}. Lastly, we can directly compare the case for $P_L=10\,\text{PW}$ with the results of \cite{Wang:2020uq}, as for all other simulations the incident power stayed in the regime of $P_L\lesssim10\,\text{PW}$ in their simulations. Here, using a laser pulse with $a_0=190$, $\lambda=1\, \mu m$ and $\tau=35\,\text{fs}$, \cite{Wang:2020uq} showed maximum photon energies of $\sim 450\,\text{MeV}$, while observing a decrease in laser to photon energy conversion efficiencies for the incident laser power $P_L \simeq 4$ PW; similar trend is also observed in Fig.\ref{fig:Thesisefficiencybeam-log} for hydrogen plasmas; see \cite{Heppe:2020ub}.

We also compare our results with theoretical predictions on maximum electron energies by \cite{Jirka:2020we}, who derived scalings on electron acceleration in the radiation dominated regime in a plasma channel. As argued by  \cite{Jirka:2020we} that radiation-reaction force can significantly alter the plasma electron dynamics, and in a best case scenario electron acceleration to ultra-high energies can occur, as also studied earlier;~\cite{Rousse:2004tc,Kumar:2006wn,Vranic:2018vz}. Thus, comparing the analytical scalings of electron acceleration by \cite{Jirka:2020we} with the inferred values of the electron energies via synchrotron radiation emission in our simulations is instructive. In our simulations,  the condition $E_{\rm field} \ll E_{\rm channel}/\mathcal{I}$ is  fulfilled. Here $\mathcal{I}$ denotes the integral of motion including the corrections caused by radiation-reaction force, as defined by \cite{Jirka:2020we},
\begin{equation}
	\mathcal{I}_0 = 1 + \left[\frac{\omega_p \mathfrak{R_0}}{2c}\right]^2
	\label{prediction1}
\end{equation}
\begin{equation}
	\mathcal{I} = \mathcal{I}_0 \left[1 + 2.3\times10^{-8}a_0^2\frac{\mathcal{I}_0}{\lambda_0[\mathrm{\mu m}]}\right]^{-1}
	\label{prediction2}
\end{equation}
$\mathcal{I}_0$ is the integral of motion without including the radiation reaction force, and $\mathfrak{R_0}$ the initial rest distance of the observed electron to the channel center. The mean energy of an electron reads as \cite{Jirka:2020we},
\noindent
\begin{equation}
	\langle \gamma_{\star}\rangle \approx 3/4\mathcal{I}^2\left(\frac{\omega_0}{\omega_p}\right)^2,
	\label{prediction3}
\end{equation}
Assuming these electros, emit photons by the well-known synchrotron emission, one can write
\begin{equation*}
	\gamma_{e, \rm max} \approx \sqrt{\frac{m_e c \varepsilon_{\gamma, \rm max}}{\hbar e B}},
\end{equation*}
where $\varepsilon_{\gamma, \rm max}$ are the maximum measured photon energies as they are denoted in Tables \ref{tab:cylindrical20} and Fig.\ref{tab:conical20} and we assumed $B\approx B_0$, which is approximately in line with the observed field strengths. To compare the predictions with the observations now we plot their ratio in Fig.\ref{fig:comparison}. 
Interestingly for low incident powers of $P=5\,\text{PW}$, our simulations yield photon energies that require electron energies to be $\sim 1.4$ times more than the analytical predictions. The divergence between the analytical scaling and the simulations results on electron acceleration become larger for increasing $P$. This suggest that additional mechanisms come into the play, \emph{i.e.} the filamentation of the laser pulse as seen in Fig.\ref{theory}. Stronger laser filamnetation of the laser pulse can  significantly reduce the bulk electron acceleration in the plasma channel. This requires a more thorough investigation in the future. 
We also recon that 2D simulations can overestimate the energies of accelerated electrons and as consequence the resulting photon energies;~\cite{Wang:2021tp}.

\section{Discussion and conclusion}\label{conclusions}

We have shown that for incident laser powers, (in the range $5$-$20\,\text{PW}$), one can efficiently convert the laser energy into the photon beams with energies up to $\sim 300\, \text{MeV}$ in carbon plasmas.  The resulting photon beams have brilliances exceeding
$10^{20}$s$^{-1}$mm$^{-1}${mrad}$^{-2}\,(0.1\,\%${BW})$^{-1}$, with a single-digit fraction of the laser to photon energy conversion efficiencies. Thus, the hitherto unexplored regime of powerful petawatt laser system $P_L \geq 10\, \text{PW}$ to generate collimated $\gamma$-ray beams appears to be promising. A major results is that the short laser pulse duration $\tau=20$ fs is preferable over longer pulse duration $\tau=40$ fs for a large range of incident laser powers. \textcolor{black}{The trend with respect to the laser to photon beam conversion efficiencies for incident laser powers is non-linear, strongly hinting at the role of the plasma instabilities associated with the laser pulse propagation in a plasma channel.}

We find that the stronger Coulomb forces due to heavier carbon ions reduces the electron acceleration and photon energies compared to the case of hydrogen plasmas. Hence, our results for C$^{6+}$-ion plasma (for $\tau = 40 \,\text{fs}$, $a_0 = 350$) are comparable to the results using  $a_0 = 190$ propagating in a hydrogen plasma channel as shown in Fig.\ref{fig:Thesisefficiencybeam-log}. Further, we show that the conical target in simulations yield higher laser to electron energy conversion efficiency, but it does not strongly improve the angular distribution of the generated photon beams and consequently their brilliances.

To summarize, we have shown the generation of  $\gamma$-rays beams with maximum brilliance exceeding $10^{20}\,\text{s}^{-1}\text{mm}^{-1}\text{mrad}^{-2}\,(0.1\,\%\text{BW})^{-1}$ for $\tau=20\,\text{fs}$ for photons with $1 \, \text{MeV}$ favoring the shorter pulse length of explicitly $20\,\text{fs}$. \textcolor{black}{We have investigated the generation of $\gamma$-photons for the laser power $P_L > 10$ PW in quantum-electrodynamics regime, where the photon generation is a stochastic effect.} For further improvement the impact of different target compositions, i.e. solid shell for the channel instead of bulk plasma, \cite{Xue:2020tz, Gu:2018uf}, might offer a way to further improve coupling between laser and plasma electrons as well as the guiding for the accelerated electrons, to improve the photon generation.

\subsubsection{Permission to Reuse and Copyright}
Figures, tables, and images will be published under a Creative Commons CC-BY licence and permission must be obtained for use of copyrighted material from other sources (including re-published/adapted/modified/partial figures and images from the internet). It is the responsibility of the authors to acquire the licenses, to follow any citation instructions requested by third-party rights holders, and cover any supplementary charges.

\subsection{Tables}

\begin{table}
\centering
\begin{tabular}{|c||c|c|c|c|}
				\hline
				$P_L$ [PW] & 5 & 10 & 15 & 20 \\ 
				\hline\hline
				$R_0$ $[\mu m]$ & 1.10 & 1.56 & 1.91 & 2.20\\
				$B_{z, max}$ $[B_0]$ & 0.85& 0.83 & 0.90 & 0.91 \\ 
				$E_{y, max}$ $[E_0]$ & 0.86 & 0.86 & 0.92 & 0.96 \\ 
				$\Tilde{N}_{\gamma, max}$ &1.00&2.15&5.41&3.06\\			
				$\varepsilon_{\gamma, max}$ [MeV] &187&203&243&275\\
				$\epsilon$ $[\%]$&0.08&2.65&0.11&0.11\\
				Brill. (@$1$ MeV)&$3.66\times10^{20}$&$1.30\times10^{20}$&$1.52\times10^{20}$&$4.00\times10^{20}$\\
				Brill. (@$10$ MeV)&$1.47\times10^{18}$&$2.20\times10^{17}$&$2.32\times10^{18}$&$4.38\times10^{18}$\\
				\hline
			\end{tabular}%
			\caption{Table of characteristic simulation results for planar target with $a_0=350$ laser pulse of duration $\tau =20$ fs,  rounded to second digit, except maximum photon energy $\varepsilon_{\gamma, \rm max}$. Maximum values depict the overall maximum of that value during the whole simulation. \textcolor{black}{$\Tilde{N}_{\gamma,\rm max}=N_{\gamma,\rm max}/N_{\gamma,\rm{max}}^{5\rm{PW}}$ is the maximum photon count during the simulation normalized to the $5\,\rm{PW}$ case.} Brilliance (Brill.) is in units of $\,\mathrm{s}^{-1}\mathrm{mm}^{-1}\mathrm{mrad}^{-2}\,(0.1\,\%\mathrm{BW})^{-1}$ with the chosen bandwidth energy corresponding to the energies referred.}
			\centering
			\label{tab:cylindrical20}
\end{table}

\begin{table}
\centering
		\begin{tabular}{|c||c|c|c|c|}
			\hline
			$P_L$ [PW] & 5 & 10 & 15 & 20 \\ 
			\hline\hline
			$R_0$ $[\mu m]$ & 1.10 & 1.56 & 1.91 & 2.20\\
			$B_{z, max}$ $[B_0]$ & 0.83& 0.85 & 1.00 & 0.99 \\ 
			$E_{y, max}$ $[E_0]$ & 0.78 & 0.89 & 0.89 & 0.93 \\ 
			$\Tilde{N}_{\gamma, max}$ &1.00&1.60&1.52&1.84\\			
			$\varepsilon_{\gamma, max}$ [MeV] &219&287&267&291\\
			$\epsilon$ $[\%]$&2.96&0.68&0.76&1.73\\
			Brill. (@$1$ MeV)&$1.91\times10^{19}$&$2.64\times10^{20}$&$4.05\times10^{19}$&$5.59\times10^{19}$\\
			Brill. (@$10$ MeV)&$1.07\times10^{17}$&$2.73\times10^{19}$&$5.98\times10^{17}$&$2.25\times10^{17}$\\
			\hline
		\end{tabular}%
	\caption{Table for quantities as in Table \ref{tab:cylindrical20} but for the laser pulse duration $\tau=40\,\mathrm{fs}$. The normalized laser pulse amplitude remain the same $a_0=350$ as in Table \ref{tab:cylindrical20}.}
	\centering
	\label{tab:cylindrical40}
	\end{table}

\begin{table}
		\centering
		 \begin{tabular}{|c||c|c|c|c|}
			\hline
			$P_L$ [PW] & 5 & 10 & 15 & 20 \\ 
			\hline\hline
			$R_0$ $[\mu m]$ & 1.10 & 1.56 & 1.91 & 2.20\\
			$B_{z, max}$ $[B_0]$ & 0.84& 0.92& 0.89 & 0.88 \\ 
			$E_{y, max}$ $[E_0]$ & 0.85 & 0.82 & 0.90& 0.93\\ 
			$\tilde{N}_{\gamma, max}$ &1.00&2.18&2.97&3.45\\
			$\varepsilon_{\gamma, max}$ [MeV] &171&208&242&256\\
			$\epsilon$ $[\%]$&4.70&10.17&6.93&8.55\\
			Brill. (@$1$ MeV)&$3.62\times10^{19}$&$6.94\times10^{19}$&$7.53\times10^{19}$&$1.41\times10^{20}$\\
			Brill. (@$10$ MeV)&$3.01\times10^{17}$&$2.73\times10^{19}$&$1.31\times10^{19}$&$4.47\times10^{17}$\\
			\hline		
			\end{tabular}
		\caption{Table of quantities for a conical target with $a_0=350$ and the laser pulse duration $\tau =20$ fs.}
		\label{tab:conical20}
	\end{table}%
	
	\begin{table}
		\centering
		\begin{tabular}{|c||c|c|c|c|}
			\hline
			$P_L$ [PW] & 5 & 10 & 15 & 20 \\ 
			\hline\hline
			$R_0$ $[\mu m]$ & 1.10 & 1.56 & 1.91 & 2.20\\
			$B_{z, max}$ $[B_0]$ & 0.88& 0.91 & 1.03 & 1.13 \\ 
			$E_{y, max}$ $[E_0]$ & 0.83 & 0.90 & 0.92 & 0.91 \\ 
			$\tilde{N}_{\gamma, max}$ &1.00&1.74&1.80&2.09\\			
			$\varepsilon_{\gamma, max}$ [MeV] &196&283&283&315\\
			$\epsilon\,[\%]$&8 $\times 10^{-4}$&5 $\times 10^{-3}$ &0.06&0.01\\
			Brill. (@$1$ MeV)&$8.16\times10^{16}$&$7.16\times10^{17}$&$8.51\times10^{18}$&$1.58\times10^{18}$\\
			Brill. (@$10$ MeV)&$2.10\times10^{14}$&$5.67\times10^{15}$&$2.16\times10^{17}$&$2.07\times10^{16}$\\
			\hline
		\end{tabular} 
		\centering
		\caption{Table of quantities for a conical target as in Table \ref{tab:conical20} but for the laser pulse duration $\tau =40$ fs.}
\label{tab:conical40}
\end{table}

\section{Nomenclature}
Two-dimensional (2D), Particle-in-cell (PIC), kilo electron volt (keV), Mega electron volt (MeV), Petawatt (PW), full-width at half-maximum (FWHM), Bandwidth (BW), Mega Tesla (MT)


\section*{Conflict of Interest Statement}

The authors declare that the research was conducted in the absence of any commercial or financial relationships that could be construed as a potential conflict of interest.

\section*{Author Contributions}

NK conceived the project and suggested to CH. CH carried out and analyzed the PIC simulations and discussed the results with NK. NK carried out the theoretical analysis of the filamentation instability. Both authors contributed to the manuscript writing.

\section*{Acknowledgments}
Results presented here are an extension of the bachelor thesis of CH, submitted to the Heidelberg University;~\cite{Heppe:2020ub}.

\section*{Supplemental Data}
Movie showing the laser propagation and interaction dynamics in a plasma channel.

\section*{Data Availability Statement}
The data presented in this manuscript can be requested to the author on a reasonable request.

\bibliographystyle{Frontiers-Harvard} 
\bibliography{LaserChannel}


\section*{Figure captions}

\setcounter{figure}{1}
\setcounter{subfigure}{0}
\begin{subfigure}
\setcounter{figure}{1}
\setcounter{subfigure}{0}
    \centering
    \begin{minipage}[b]{0.5\textwidth}
    \centering
        \includegraphics[width=1.3\linewidth]{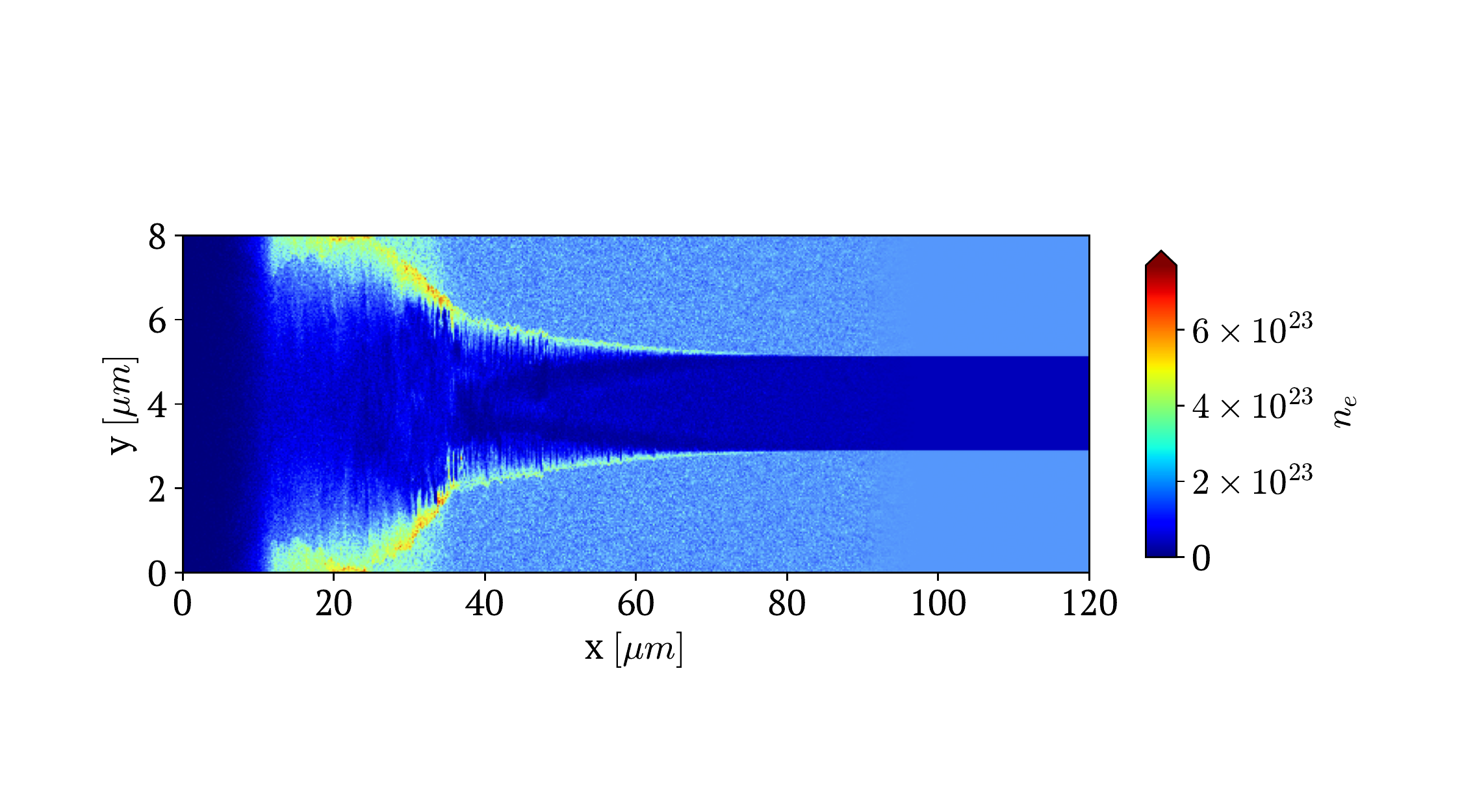}
        \caption{Plasma density, $n_e$}
        \label{fig:1a}
    \end{minipage}  
   
\setcounter{figure}{1}
\setcounter{subfigure}{1}
        \begin{minipage}[b]{0.5\textwidth}
        \centering
        \includegraphics[width=1.3\linewidth]{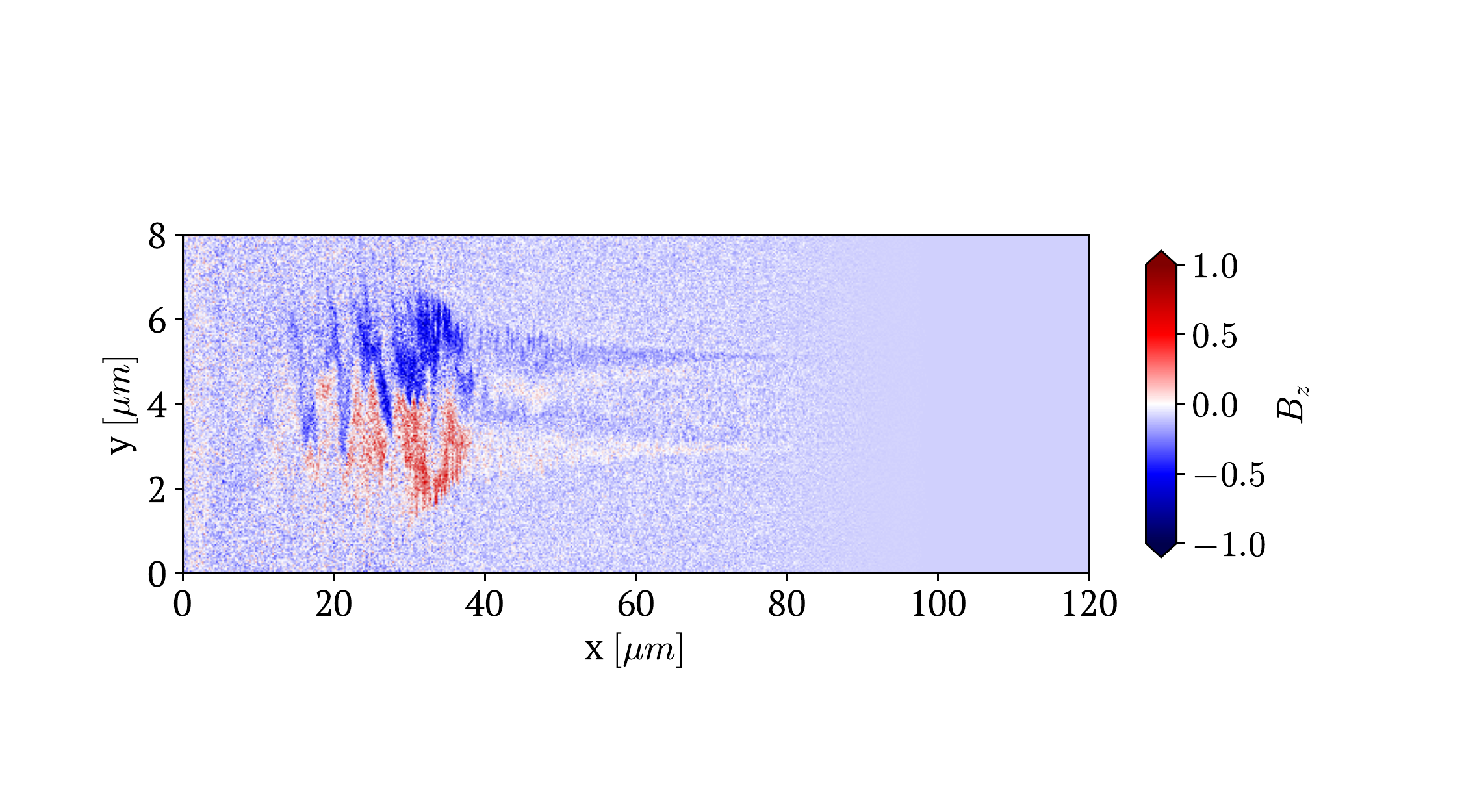}
        \caption{Azimuthal magnetic field, $B_z$}
        \label{fig:1b}
    \end{minipage}
    
\setcounter{figure}{1}
\setcounter{subfigure}{2}
    \begin{minipage}[b]{0.5\textwidth}
       \centering
        \includegraphics[width=1.4\linewidth]{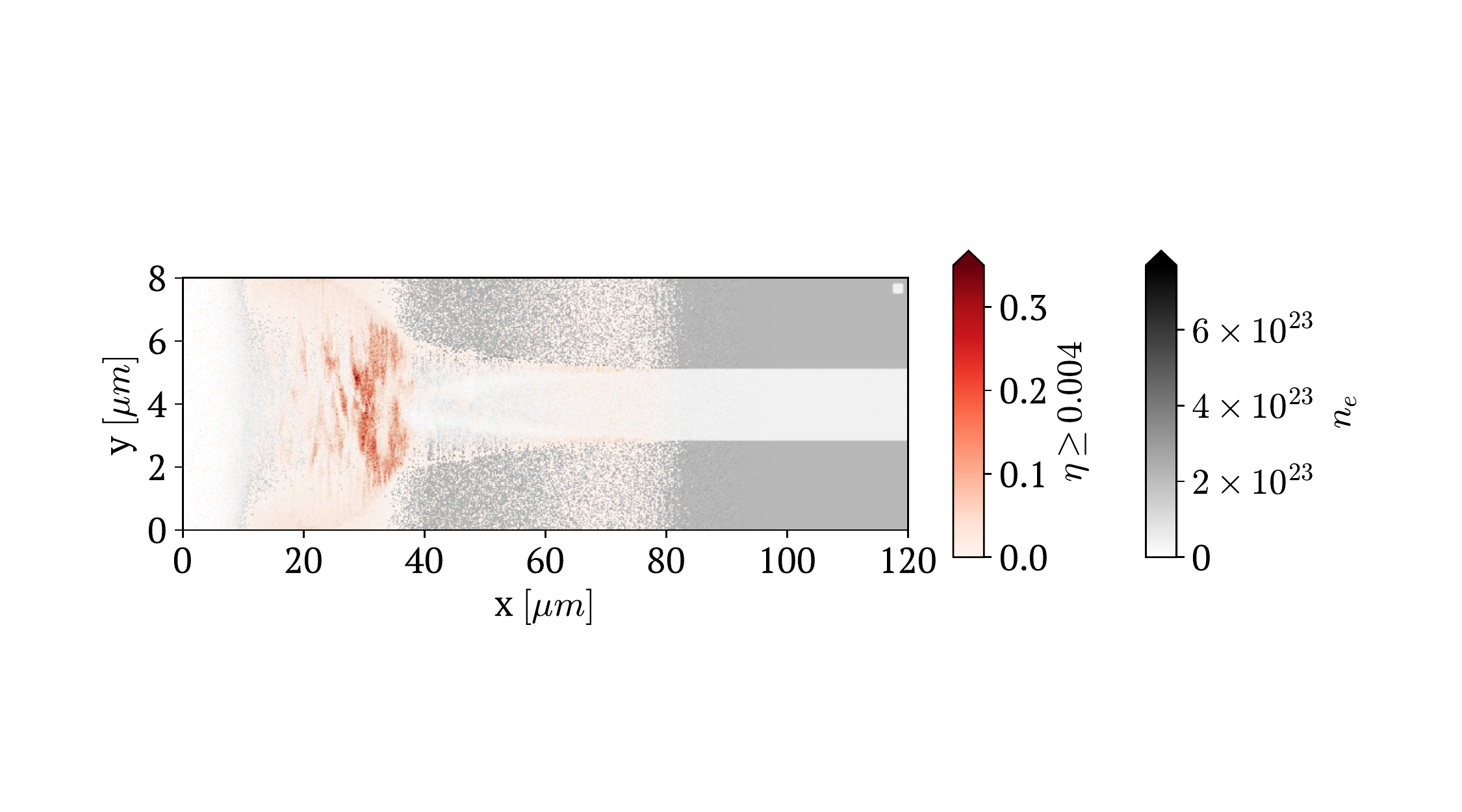}
        \caption{$\eta \geq 0.0004$ (red), $n_e$ [$n_{cr}$] (grey)}
        \label{fig:1c}
    \end{minipage}

\setcounter{figure}{1}
\setcounter{subfigure}{-1}
    \caption{Electron density in units of $n_{cr}$, azimuthal magnetic field strength $B_z$ in units of $B_0$ of the laser field and electron density (grey) overlayed with emission parameter $\eta \geq 0.0004$ (Fig.\ref{PowSynch}, red) for a simulation with a laser pulse $P_L=5\,\text{PW}$, $\tau=40\,\text{fs}$ and $a_0=350$ hitting a planar target of $C^{6+}$ plasma at around $t\approx 1400\,\text{fs}$.}
    \label{fig:1}
\end{subfigure}

\begin{figure}
    \centering
	\includegraphics[width=0.8\linewidth]{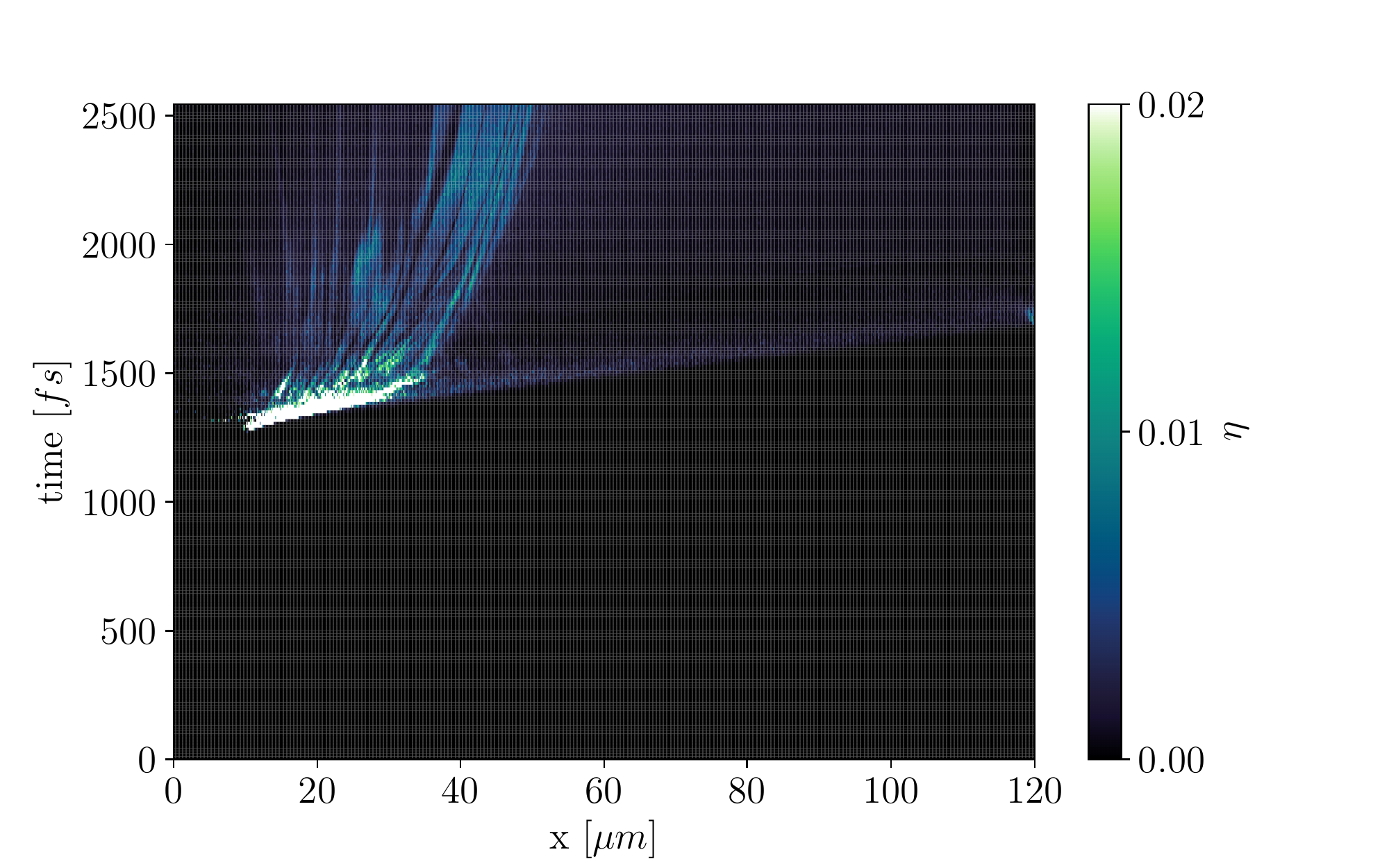}
	\caption{Synchrotron emission parameter $\eta$ along the symmetry axis ($y= 4 \,\mathrm{\mu m }$) of the plasma channel over time for simulations of a $\tau = 40 \,\text{fs}$ laser pulse with incident Power $P_L = 5 \,\text{PW}$ and a normalized amplitude $a_0=350$. The colorbar indicates the Synchrotron emission power parameter $\eta$ (\ref{PowSynch}) and is limited to values up to $\eta  _{max}= 0.005$ (higher values are depicted as $\eta _{max}$) to enhance visibility.}
	\label{fig:Eta}
\end{figure}

\begin{figure}
    \centering
	\includegraphics[width=0.8\linewidth]{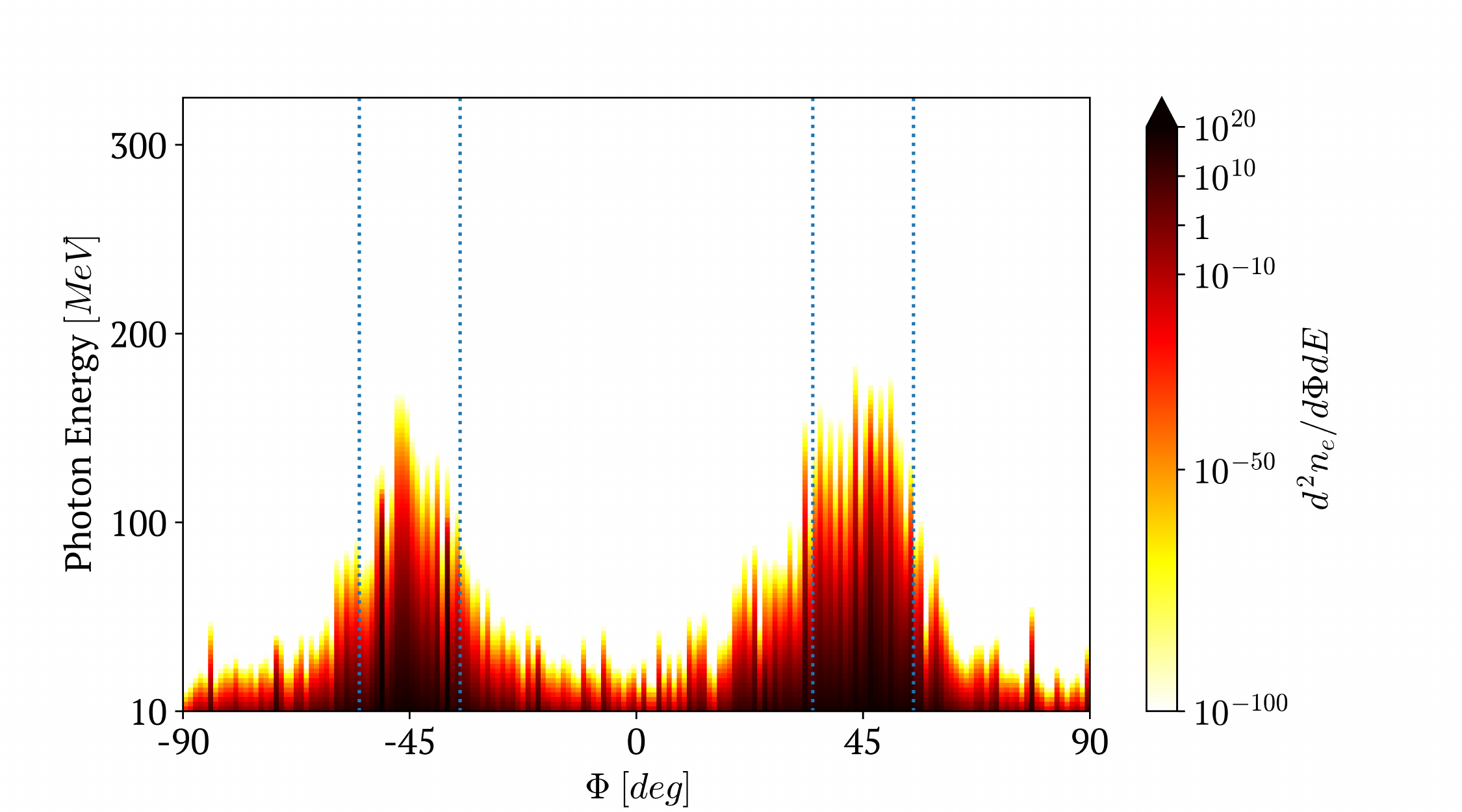}
	\caption{Angular distribution of photon energy spectrum in degree with respect to the channel’s symmetric axis in  MeV for a simulation with a laser pulse $P_L=5\,\text{PW}$, $\tau=40\,\text{fs}$  and $a_0=350$ hitting a cylindrical target of $C^{6+}$ plasma, averaged over the duration of the simulation. The dotted blue line indicates the beams defined as the emission cone at $\Phi =\pm 45^{\circ}\pm 10^{\circ}$.}
	\label{fig:Angular}
\end{figure}

\begin{figure}
\centering
	\includegraphics[width=0.7\linewidth]{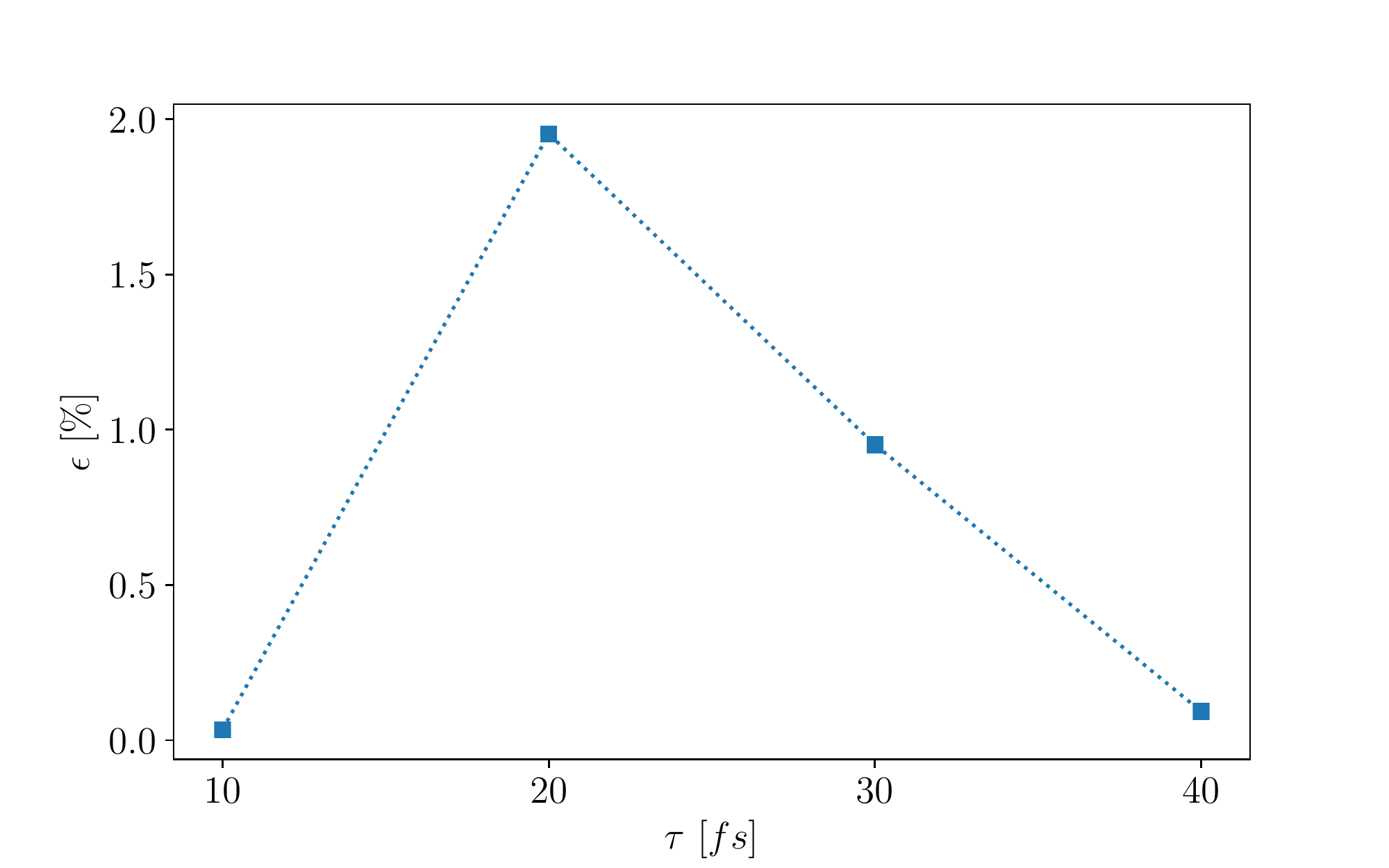}
	\caption{$\tau$-scan for a $10\, {\rm PW}$ laser pulse with $a_0=350$ regarding conversion efficiency into a collimated a high energy $\gamma$-ray beam ($\geq 10\,\text{MeV}$) with beam width $\Delta\theta=20^{\circ}$ in percentages for a cylindrical $C^{6+}$-ion plasma target.}
	\label{fig:tau-scan}
\end{figure}

\setcounter{figure}{5}
\setcounter{subfigure}{0}

\begin{subfigure}
\setcounter{figure}{5}
\setcounter{subfigure}{0}
\centering
\begin{minipage}[b]{0.5\textwidth}
		\includegraphics[width=\linewidth]{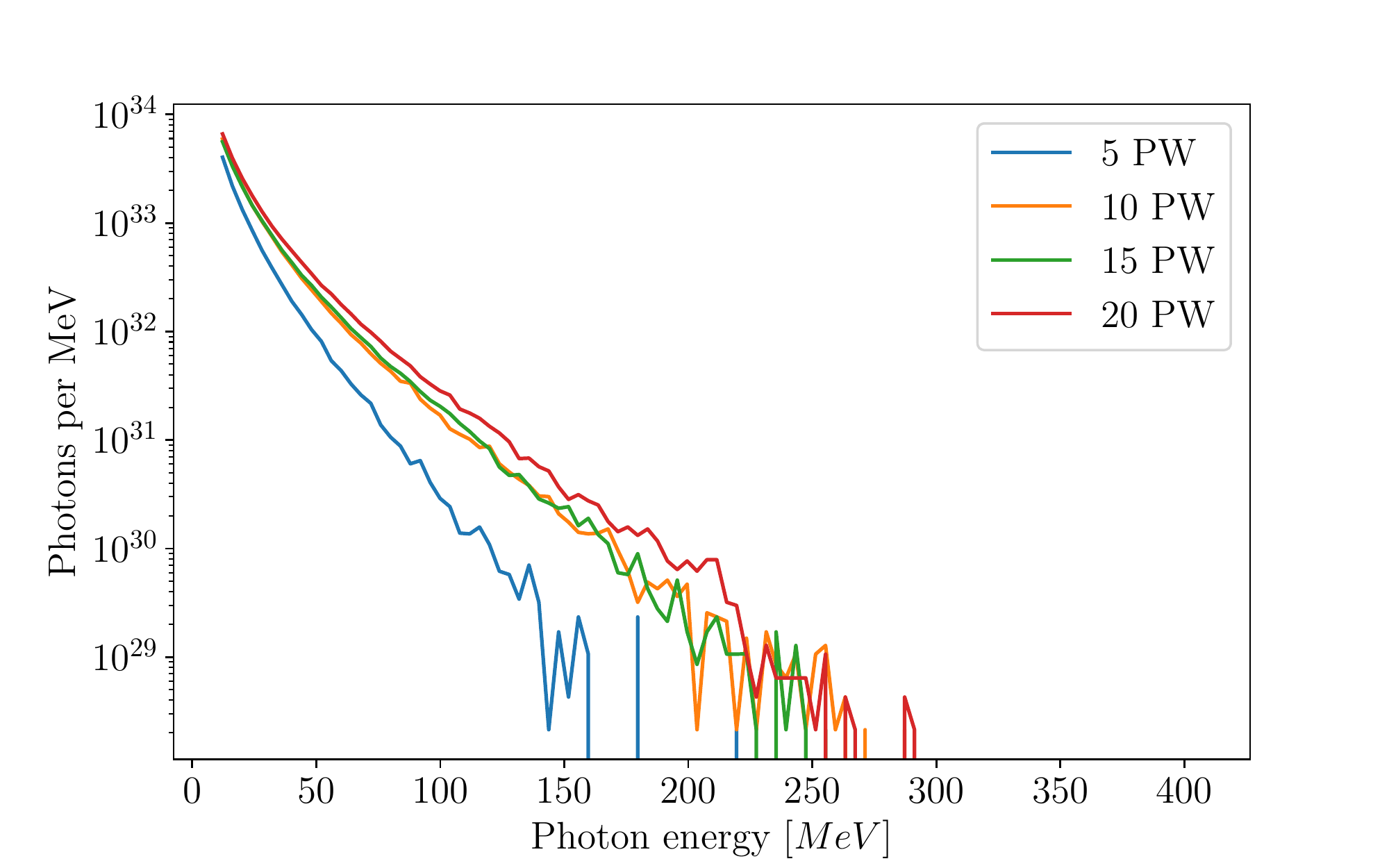}
		\caption{$\tau=20\,\text{fs}$}
		\end{minipage}  
\setcounter{figure}{5}
\setcounter{subfigure}{1}
    \begin{minipage}[b]{0.5\textwidth}
	
		\includegraphics[width=\linewidth]{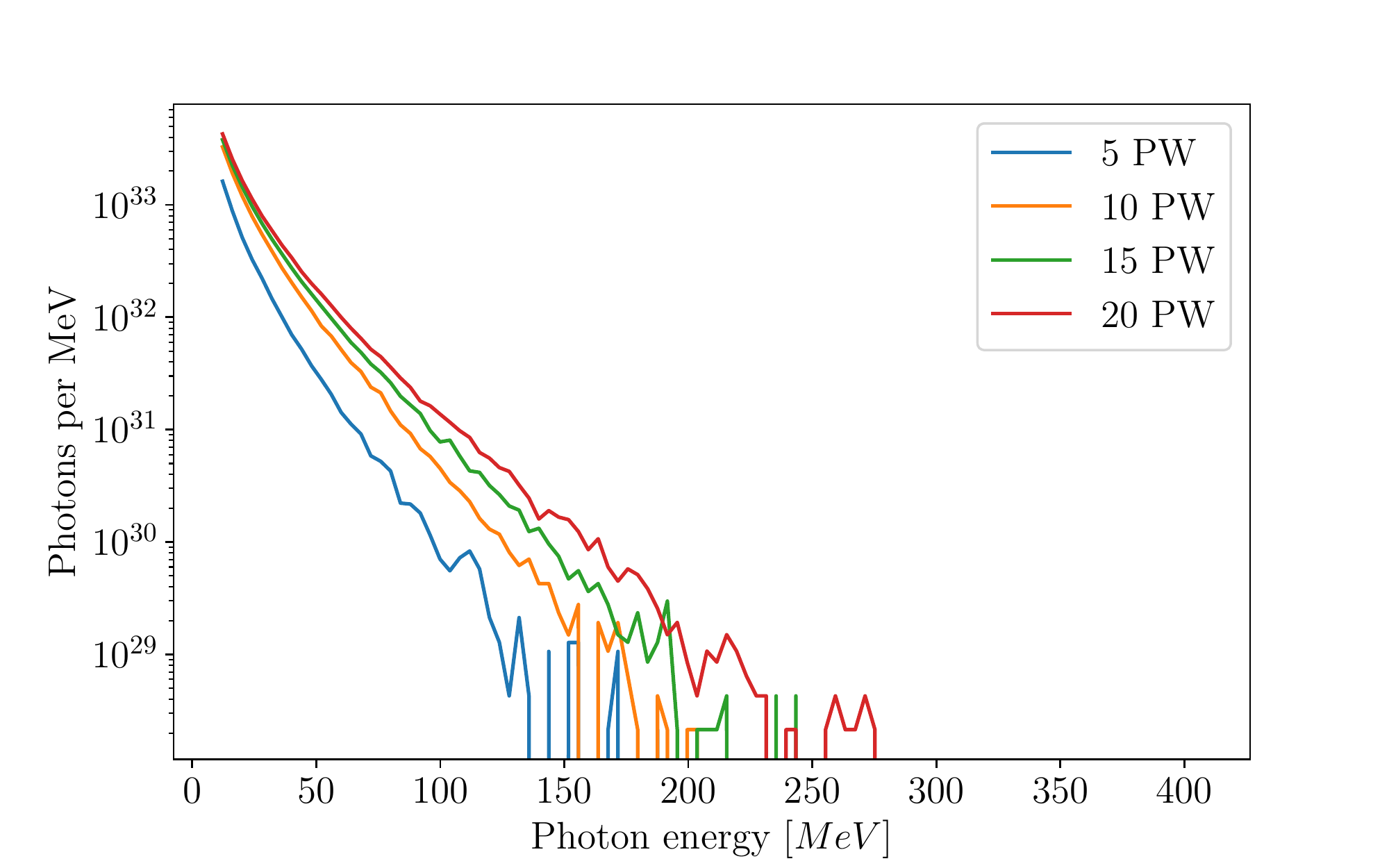}
		\caption{$\tau=40\,\text{fs}$}
	\end{minipage}
	
\setcounter{figure}{5}
\setcounter{subfigure}{-1}

		\caption{Photon emission spectra for photons with energies $\geq 5\,\text{MeV}$, for a $a_0=350$ laser impacting a planar target C$^{6+}$-plasma target.}
	\label{fig:Spectrum-cylindric}
\end{subfigure}

\setcounter{figure}{6}
\setcounter{subfigure}{0}

\begin{subfigure}
\setcounter{figure}{6}
\setcounter{subfigure}{0}
\centering
\begin{minipage}[b]{0.5\textwidth}
		\includegraphics[width=\linewidth]{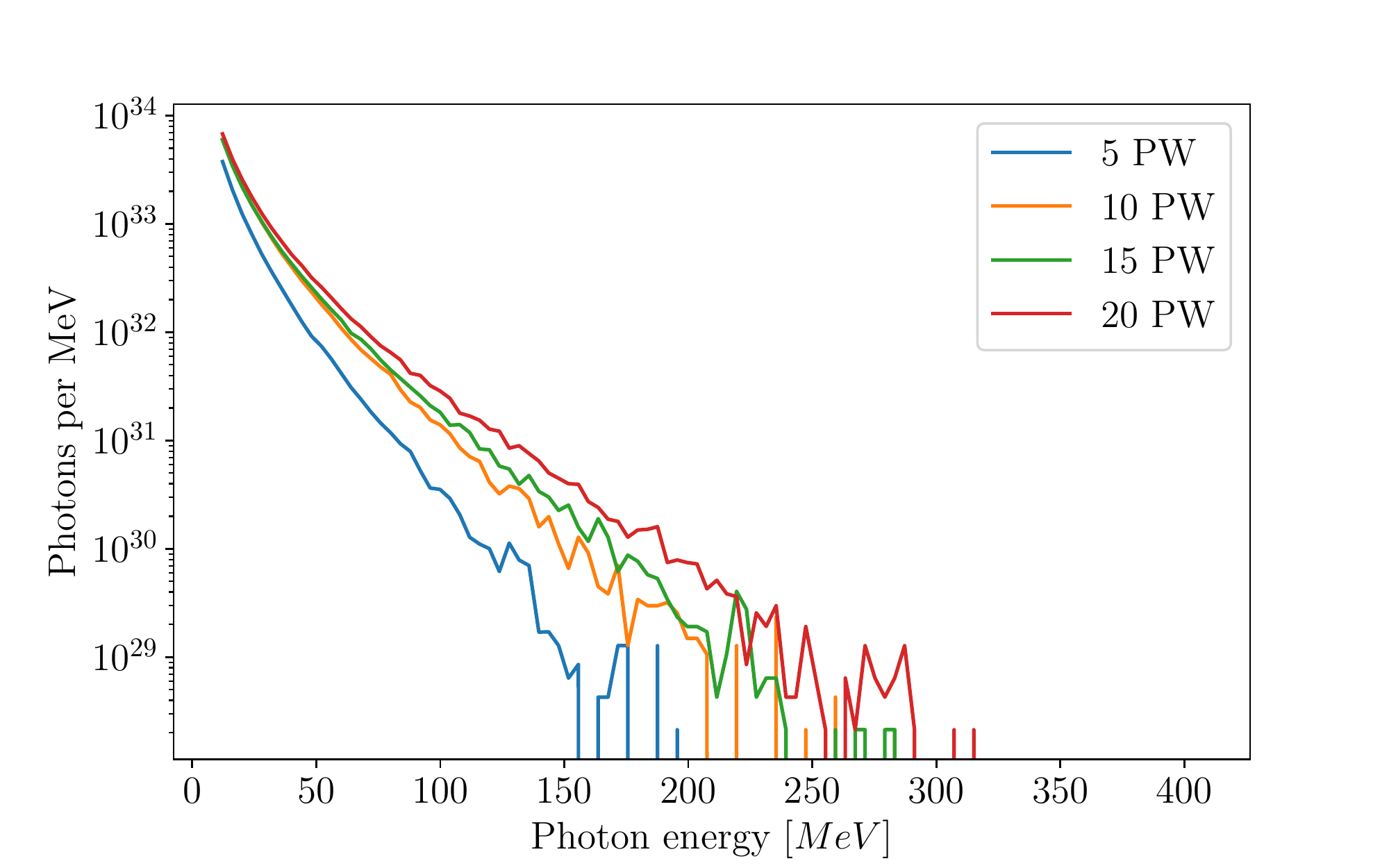}
		\caption{$\tau=20\,\text{fs}$}
		\end{minipage}  
\setcounter{figure}{6}
\setcounter{subfigure}{1}
    \begin{minipage}[b]{0.5\textwidth}
	
		\includegraphics[width=\linewidth]{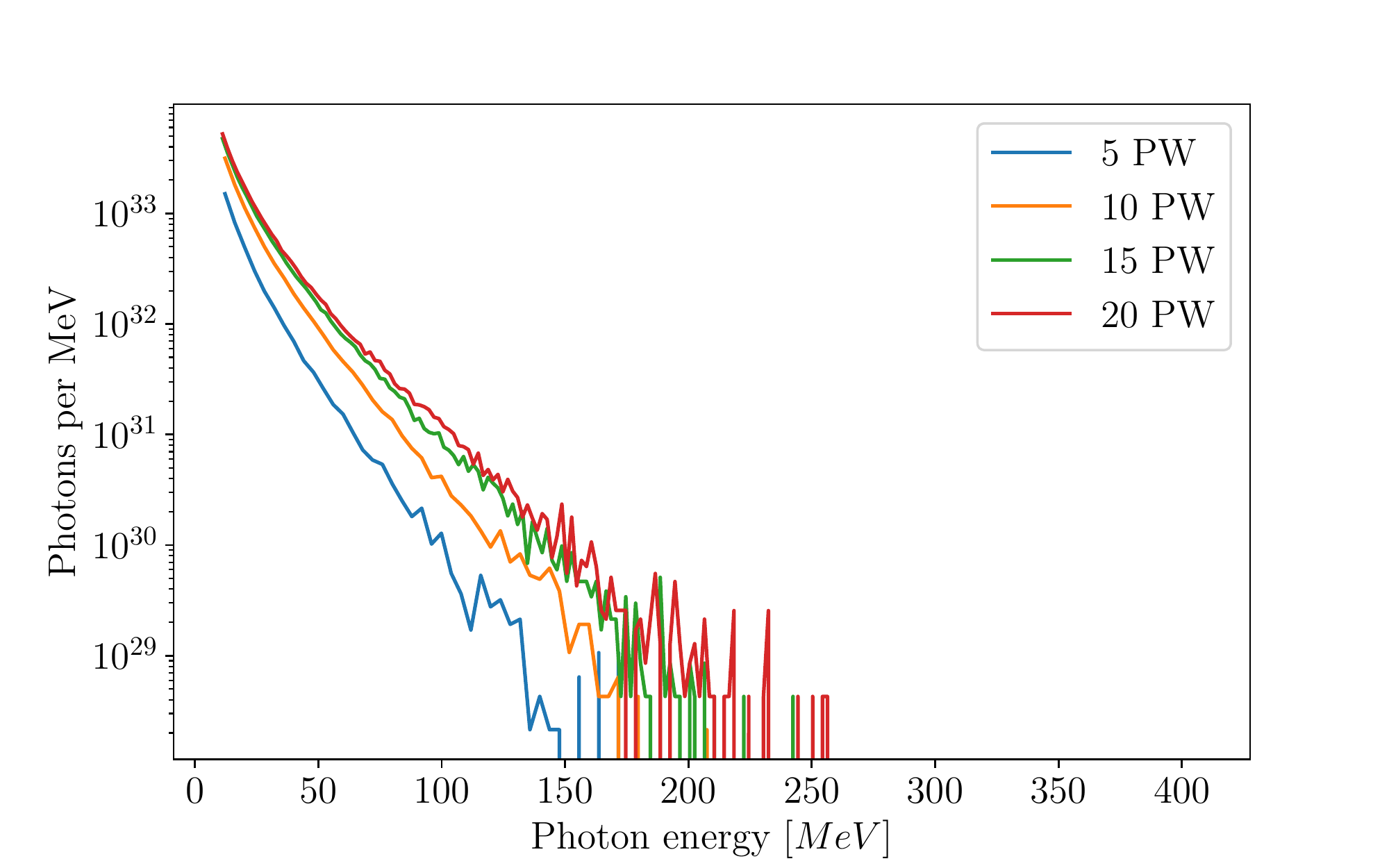}
		\caption{$\tau=40\,\text{fs}$}
	\end{minipage}
	
\setcounter{figure}{6}
\setcounter{subfigure}{-1}

		\caption{Photon emission spectra for photons with energies $\geq 5\,\text{MeV}$, for a $a_0=350$ laser impacting a conical $C^{6+}$-plasma target.}
	\label{fig:Spectrum-conical}
\end{subfigure}

\begin{figure}
\centering
	\includegraphics[width=0.7\linewidth]{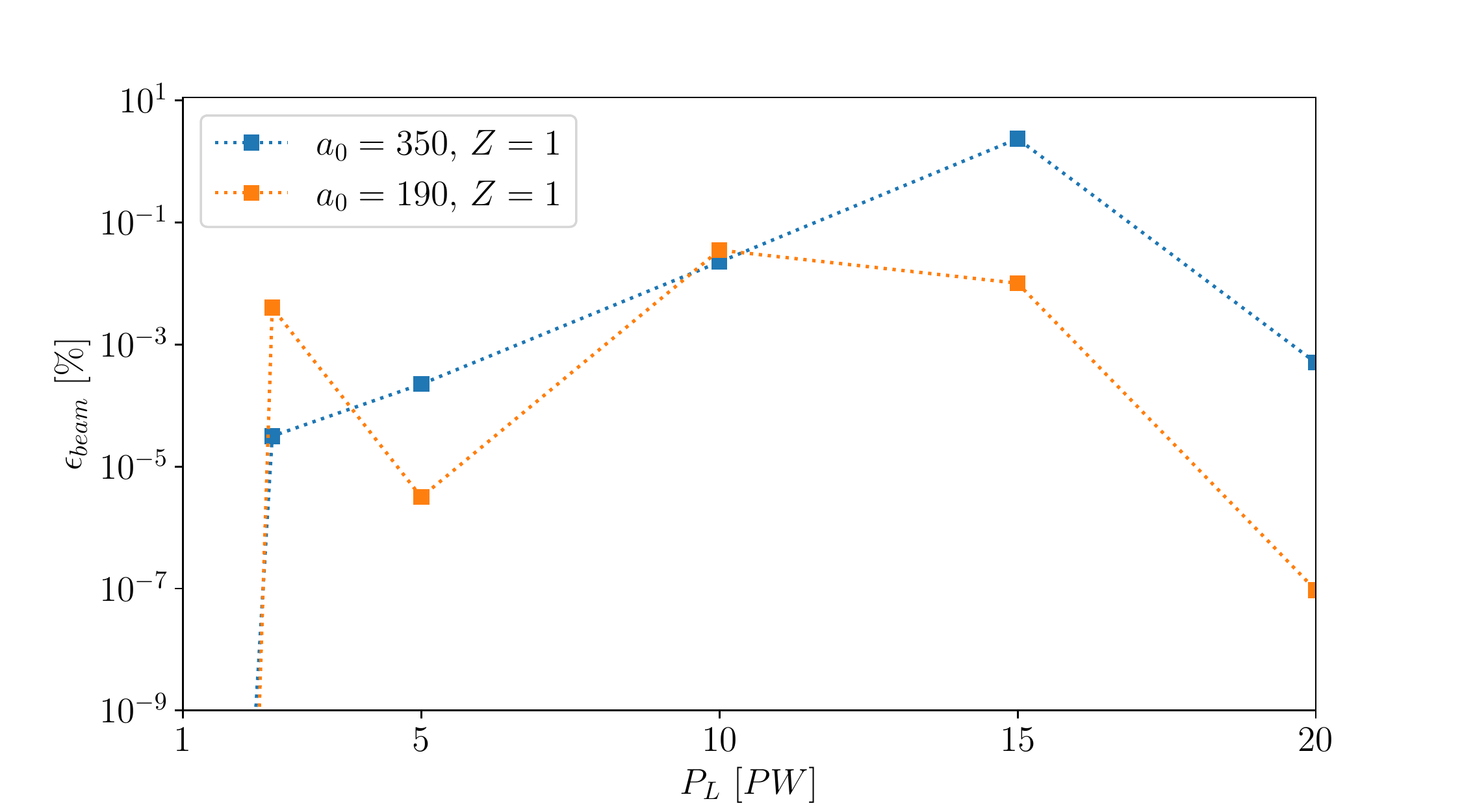}
	\caption{Conversion efficiency of laser energy ($\tau = 40\,\text{fs}$) impacting a $Z=1$ $H^{+}$-plasma into a high energy $\gamma$-ray beam ($\geq 10\,\text{MeV}$) with a beam width of $\Delta\Theta = 20^\circ$ in percentages of the total energy for simulations carried out for \cite{Heppe:2020ub}.}
	\label{fig:Thesisefficiencybeam-log}
\end{figure}

\begin{figure}
   \centering
	\includegraphics[width=0.8\linewidth]{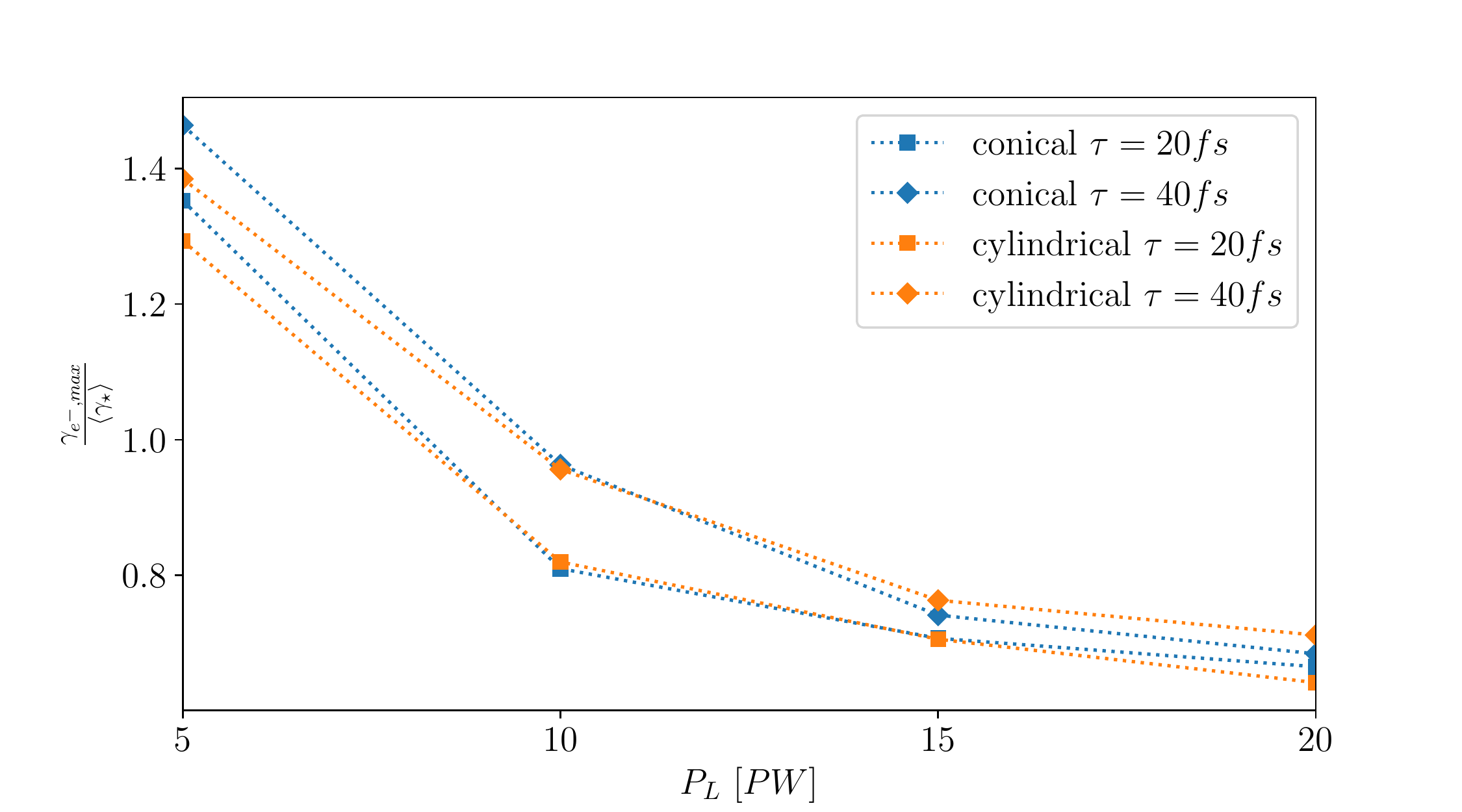}
	\caption{Ratio of predicted electron Lorentz-factor $\gamma_{\star, \rm max}$ using Eq.\eqref{prediction3} to $\gamma_{e, \rm max}$ the observed/measured Lorentz-factor as function of incident power $P_L$}
	\label{fig:comparison}
\end{figure}

   



\end{document}